\documentclass{article}  

\usepackage{graphicx}
\usepackage{amsthm}
\usepackage{amsmath,amssymb}
\usepackage{graphicx}
\usepackage{subcaption}
\usepackage{enumitem} 
\usepackage{mathtools}
\usepackage{enumitem}
\usepackage{algorithm}
\usepackage{textcomp}
\usepackage{siunitx}

\newtheorem{Definition}{Definition}

\newtheorem{Proposition}{Proposition}
\newtheorem{Lemma}{Lemma}
\newtheorem{Corollary}{Corollary}

\newtheorem{customProposition}{Proposition}

\newtheorem{customLemma}{Lemma}

\newcommand{\LQRTrack}[1]{u_\mathrm{LQR}^{#1}}
\newcommand{\LQRSwitch}[1]{u_\mathrm{switch}^{#1}}
\newcommand{\USwitch}[2]{u_\mathrm{switch}^{#1}}
\newcommand{\Sol}[2]{\Sigma(#1, #2)}
\newcommand{\Cost}[3]{\mathrm{cost}_{#1}(#2, #3)}

\newcommand\norm[1]{\left\lVert#1\right\rVert}

\begin{document}

\title{Computation of Feedback Control Laws Based on Switched Tracking of Demonstrations\thanks{This work as supported by the project GA21-09458S of the Czech Science Foundation GA \v{C}R and institutional support RVO:67985807.}}

\author{Ji{\v r}{\'\i} Fejlek$^{1,2}$\footnote{ORCID: 0000-0002-9498-3460} \;and Stefan Ratschan$^{1}$\footnote{ORCID: 0000-0003-1710-1513}\\ 
$^1$The Czech Academy of Sciences, Institute of Computer Science,\\ Pod Vod\'{a}renskou v\v{e}\v{z}\'{i} 271/2, Prague, 182 07, Czech Republic\\
$^2$Czech Technical University in Prague, Faculty of Nuclear Sciences and Physical Engineering, \\Department of Mathematics, Trojanova 13, Prague, 120 00, Czech Republic}

\maketitle
\thispagestyle{empty}
\pagestyle{empty}

\begin{abstract}                          
A common approach in robotics is to learn tasks by generalizing from special cases given by a so-called demonstrator. In this paper, we apply this paradigm and present an algorithm that uses a demonstrator (typically given by a trajectory optimizer) to automatically synthesize  feedback controllers for steering a system described by ordinary differential equations into a goal set.  The~resulting feedback control law switches between the demonstrations that it uses as reference trajectories. In comparison to the direct use of trajectory optimization as a control law, for example, in the form of model predictive control, this allows for a~much simpler and more efficient implementation of the controller. The synthesis algorithm  comes with rigorous convergence and optimality results, and computational experiments confirm its efficiency.
\end{abstract}

\section{Introduction}

In this paper, we provide an~algorithm that~for a given system of ordinary differential equations with control inputs synthesizes a controller that steers the system from a given set of initial set of states to a given goal set. The construction is based on the notion of a demonstrator~\cite{Rav:19}, for example, in the form of a trajectory optimization method. Such a demonstrator is a procedure that is already able to provide control inputs that steer the system into the goal set. However, we assume that the demonstrator itself is unsuitable for direct (e.g., online) usage, for example, due to lack of efficiency. Thus the produced control inputs take the role of demonstrations~\cite{demonstration1,demonstration2,learningCLF1} that we use to learn a feedback control law offline from. Assuming that~some cost is assigned to~control, the resulting control law also inherits the demonstrator's performance in terms of this cost.

The offline synthesis algorithm constructs the desired controller using a loop that (1) learns a control law, generalizing the current demonstrations to the whole statespace, (2) searches for a counterexample to the desired properties of this control law, and (3) queries the demonstrator for a new demonstration from this counterexample. It iterates this loop until the result is good enough. During this process, it maintains a certificate to reduce the simulation time needed in counterexample search. The algorithm extends construction of control laws based on demonstrations~\cite{demonstration1,learningCLF1,lib2,lib3}, namely LQR-trees~\cite{LQRtrees2,LQRtrees1}, and learning certificates of system behaviour from data/demonstrations~\cite{learningCLF3,Rav:19,neural,learningCLF4,learningCLF5,sim1}.

We prove that under some mild assumptions, finitely many cycles of this loop generate a controller that steers the system into the goal set. This is a significantly stronger result than in~\cite{LQRtrees2,LQRtrees1}, which only describes the behaviour of the algorithm for the number of iterations tending to infinity. Moreover, we prove that the generated controllers asymptotically reach the performance of the demonstrator.

We also do computational experiments on several examples of dimension up to twelve that~demonstrate the~practical applicability of the~method. We compare our algorithm with controller synthesis fully based on~system simulations~\cite{LQRtrees2}. In this comparison, our synthesis algorithm runs significantly faster (between 50\% and 95\%, which corresponds to reducing several hours of computation time to several minutes in the most extreme cases), while producing controllers of similar performance.

The~structure of the~paper is as follows. In~Section~\ref{sec:problem}, we state the~precise problem and discuss related work. In~Section \ref{sec:generic}, we introduce the~general layout of our algorithm. Section~\ref{sec:track} is the core of the paper, developing the details of the synthesized controllers and the synthesis algorithm, exploring their theoretical properties. In Section~\ref{sec:implementation}, where we describe implementation of the synthesis algorithm. In Section~\ref{sec:Examples} we provide computational experiments. And finally, Section~\ref{sec:conclusion} contains a~conclusion. Proofs of all propositions are included in an~appendix.

\section{Problem Statement and Related Work}
\label{sec:problem}

Consider a~control system of the~form
\begin{equation}
\label{system}
\dot{x} = F(x,u) 
\end{equation}
where $F\colon\mathbb{R}^n\times \mathbb{R}^m\mapsto\mathbb{R}^n$ is a~smooth function. Let $U \subset \mathbb{R}^m$ be a bounded set of control inputs and let $T>0$. For any initial point $x_0$ and a control input $u\colon[0, T]\mapsto U$, we denote the corresponding solution of~(\ref{system}) by $\Sol{x_0}{u}$, which~is a~function in $[0, T]\mapsto \mathbb{R}^n$. We also use the~same notation to~denote the~solution of the~closed loop system resulting from a feedback control law $u\colon \mathbb{R}^n\times[0, T]\mapsto\mathbb{R}^m$.

Now consider a~compact set of initial states $I\subset \mathbb{R}^n$ and a~compact set of goal states $G \subset \mathbb{R}^n$ such that $G\subset I$. We wish to construct a~control law $u$ such that $\Sol{x_0}{u}$ reaches $G$ for all $x_0 \in I\setminus G$. 

Assume that we already have a procedure that provides the necessary control input $u$ for each $x_0 \in I\setminus G$. Such a procedure could, for example, be based on trajectory optimization~\cite{Bet:10}, path planning algorithms~\cite{RRT3}, model predictive control (MPC)~\cite{MPC1}, or even a human expert~\cite{learningCLF3}. But we also assume that this procedure is for some reasons (computational complexity, time constraints, unreliability in its implementation, etc.) unsuitable for direct (e.g., online) usage. Therefore we will use the procedure as a so-called demonstrator~\cite{Rav:19,demonstration1,learningCLF1}, that is, a black-box that generates the desired control inputs in form of exemplary system trajectories, and from which we will learn a control law suitable for direct use. 

We also assume that this demonstrator achieves certain  performance, that the control law that we will learn, should preserve. We measure this 
performance of a demonstration of length $T$ in terms of a cost functional in the integral form
\begin{equation}
\label{cost}
V_T(x,u) := \int_0^{T} P(x(t),u(t),t) \;\mathrm{d}t,
\end{equation}
where $P\colon\mathbb{R}^n\times \mathbb{R}^m\times \mathbb{R}\mapsto\mathbb{R}$ is a~smooth function. We denote by $\Cost{V}{x_0}{u}$ the~cost of the~solution $\Sol{x_0}{u}$ and its corresponding control input (i.e., $u$ itself, if it is an~open loop control input, and $u'\colon [0,T]\rightarrow \mathbb{R}^m$ with $u'(t)=u(\Sol{x_0}{u}(t), t)$, if $u$ is a~feedback control law). 

To sum up, in this paper we contribute an algorithm that uses a given demonstrator to learn a simpler feedback control law $u$ that inherits the following key properties of the demonstrator:
\begin{enumerate}
\item $\Sol{x_0}{u}$ reaches $G$ for all $x_0 \in I$ (\emph{reachability})
\item $u$ achieves performance of the demonstrator in terms of $V_T(x,u)$ (\emph{relative optimality})
\end{enumerate}
Internally, the algorithm will construct a certificate supporting the reachability property. It uses this certificate to learn information about the global behavior of the learned control law. This information may also be useful for independent verification after termination. We will refer to such a certificate, that we will formally define later, as a \emph{reachability certificate}.

We will now discuss the contribution over related work in more detail. Learning from demonstrations  is a popular approach to control design in robotics~\cite{demonstration1,learningCLF1}. However, such approaches are usually not fully algorithmic, assuming the presence of an intelligent human demonstrator. 

Sampling-based planning algorithms~\cite{RRT7} are completely automatized but usually ignore constraints in the form of differential equations, and simply assume that it is possible to straightforwardly steer the system from any state to any further state that is close. 


The most similar approach to the construction of control laws that we present here is based on LQR-trees~\cite{LQRtrees2,LQRtrees1} that systematically explores trees of trajectories around which it builds controllers using linear-quadratic regulator (LQR) based tracking. Our work improves upon this in two main aspects: First, we explore the~theoretical properties of the~resulting algorithm more deeply
and hence obtain stronger results for successful construction of a control law concerning convergence and optimality of the result. 
Second, those methods ensure correctness of the constructed control law separately around each demonstration, whereas we construct a global certificate in parallel to the control law. This avoids the need for simulating whole system trajectories to check reachability~\cite{lib3,lib4,lib6,LQRtrees2}, and for computing a separate local certificate for each individual trajectory tracking system~\cite{LQRtrees1} via construction of control funnels~\cite{funnel,funnels5}, which are highly non-trivial to compute~\cite{SOS_funnels,Fejlek:21}. Our computational experiments confirm the advantage of such an approach.


An alternative to learning controllers in the form of LQR tracking would be to represent the learned control law using neural networks~\cite{GPS1,GPS2,Mordatch:14,learn_MPC,learn_MPC2}. The precise behaviour of neural networks is however difficult to~understand and verify. In contrast to that, our algorithms provide rigorous convergence proofs.

Yet another related approach is to construct control laws in the form of control Lyapunov functions (CLF)~\cite{CLF1} that---under certain conditions---represent a~feedback control law. Learning CLFs from data/ demonstrations \cite{learningCLF3,Rav:19,neural,learningCLF4,learningCLF5} improves upon the scalability issues of computing CLFs using dynamic programming~\cite{Ber:05} or sum-of-squares (SOS) relaxation~\cite{SOS1}. 
Our algorithm indeed computes its reachability certificate in a similar way as algorithms for learning CLFs~\cite{Rav:19}.  However, by also learning a~separate control law directly, it obtains control laws that mimic the demonstrations, inheriting  performance of the demonstrator. In contrast to that, the feedback control law represented by a CLF is only inverse optimal---optimal for \emph{some} cost functional, but not necessarily for the desired one~\cite{inverse1}. 

The underlying algorithmic principle of a learning and testing loop is becoming increasingly important. Especially, it is used in the area of program synthesis under the term counterexample guided inductive synthesis (CEGIS)~\cite{Solar:08}. The systematic search for counterexamples to a certain desired system property is also an active area of research in the area of computer aided verification where it goes under the name of \emph{falsification}~\cite{Yashwant:11}, that we will also use here.



\section{Generic Algorithm}
\label{sec:generic}

In this section, we give a~brief informal version of our algorithm that is shown as Algorithm~\ref{alg:generic} below. In the following sections we will then make it precise and prove properties of its behavior.

\begin{algorithm}[htb]
  \caption{Generic Controller Synthesis}
  \label{alg:generic}
  \begin{description}
\item[Input:] A~control system $\dot{x} = F(x,u)$ 
\item[Output:] A~control law that steers the~given control system from an initial set $I$ to a goal set $G$ 
  \begin{enumerate}
\item Generate an initial set of demonstrations  $\mathcal{T}$
\item Iterate:
\begin{enumerate}
\item Learn a~reachability certificate candidate $L$ using the~demonstrations in~$\mathcal{T}$. 
\item Construct a~feedback control law $u$ using the~current set of demonstrations  $\mathcal{T}$. 
\item Use the candidate $L$ to test whether the~control law $u$ meets the desired reachability property. If a~counterexample $x$, from where $L$ does not certify reachability, is found, generate a~demonstration starting from $x$ and add it into $\mathcal{T}$. 
\end{enumerate}
\end{enumerate}
\end{description}
\end{algorithm}

The algorithm uses a learning-and-falsification loop. Within the loop it learns a candidate for a reachability certificate, and a corresponding feedback control law. 
A learned reachability certificate candidate will satisfy certain Lyapunov-like conditions that we will define in the next section. The algorithm then constructs a~feedback control law from the available demonstrations and uses  the~candidate  to check if this control law steers the~system into $G$. 
If it finds a counterexample, a state in which  eeachability certificate candidate does not certify reachability, then the algorithm generates a~new demonstration from the~point where the~problem was detected, which results in a new feedback control law. This procedure repeats until no new counterexamples are found.

\section{Algorithm with Tracking Control}
\label{sec:track}

In this section, we adapt the~generic algorithm to a~concrete instantiation based on tracking. For this, we will define a notion of reachability certificate that will allow us to~check reachability of the~generated control law and to~identify counterexamples. These counterexamples are used as initial points for new demonstrations. Most of the theoretical results described here will hold for general tracking, but we focus on a~control law that is constructed using time-varying linear-quadratic regulator (LQR) models along the~generated demonstrations. 

We will develop the algorithms as follows. First, in Section~\ref{sec:demonstrator} we  formalize the used notion of a demonstrator.  Then, in Section~\ref{sec:control-switch}, we define a switching controller based on tracking, and a corresponding notion of reachability certificate. In Section~\ref{sec:algorithm}, we provide the~algorithm. In Section~\ref{sec:properties} we discuss properties of the~algorithm with an LQR tracking controller.

\subsection{The Demonstrator}
\label{sec:demonstrator}

A~key part of the~algorithm is the~set of trajectories provided by the~demonstrator. They will have the following form:
\begin{Definition}
\label{demonstration_def}
Let $X \subset \mathbb{R}^n$ and $U \subset \mathbb{R}^m$ be bounded sets and let $T > 0$. A~\emph{demonstration from $x_0 \in X$} is a~trajectory $(x,u)$ with $x\colon [0, T] \mapsto X$ and $u\colon [0, T]\mapsto U$, both continuous, that~satisfies the control system~(\ref{system}), and for which $x(0)=x_0$ and $x(T)\in G$.  
\end{Definition}

Note that we assume a fixed length $T$ of all demonstrations. It would not be difficult to generalize the presented results to demonstrations
of variable length. However, adhering to demonstrations of fixed length simplifies our presentation.

We obtain such demonstrations from a demonstrator. It would be possible to use the demonstrator itself as a~feedback control law that periodically computes a~demonstration from the~current point and applies the~corresponding control input for a~certain time period. However, this has two disadvantages: First, the~run-time of the~demonstrator might be too long for real-time usage. Second, the~control law provided by the~demonstrator is often defined implicitly, using numerical optimization, which~makes further analysis of the~behavior (e.g. formal verification) of the~resulting closed loop system difficult. Hence, we use demonstrations to define a new control law that is easier compute and analyze.

We formalize a demonstrator as follows:
\begin{Definition}
\label{demonstrator_def}
A \emph{demonstrator} on a set  $W \subseteq X$ is a procedure that assigns to each system state $x_0 \in W$ a demonstration $(x, u)$ with $x(0)=x_0$. 
\end{Definition}
In addition, we assume that the~demonstrator somehow takes into account the~cost functional \eqref{cost}, although the~result does not necessarily have to~be optimal. 

In practice, we can generate only finitely many demonstrations. And thus, we need to~determine when the~set of demonstrations is sufficient. The~minimum requirement on the~set of demonstrations is reachability of the~system with a controller based on these demonstrations. And this is the~requirement that~we will use for our algorithm.

\subsection{Switching Control}
\label{sec:control-switch}

Now we will construct a~feedback control law that combines individual tracking controllers in such a way that the it steers the system into the goal set. In~\cite{LQRtrees1}, multiple certificates for each separate demonstration so called \emph{funnels} are used to construct a feedback control law. Computing these is however costly and it is required to cover the whole initial set $I$ with these funnels. In~\cite{LQRtrees2}, the exact certificates are dropped in favor of merely estimating funnels using simulations of the whole system trajectories from randomly generated samples. 

To avoid computing multiple certificates and the issue of covering $I$ with them, we introduce one single reachability certificate and switching of demonstrations. Instead of just following a~single demonstration to the goal, we will follow a~given demonstration only for some time period, then re-evaluate whether the~current target demonstration is still the~most convenient one, and switch to a~different one, if it is not. Consequently, we can check reachability from any given point $x(0)$ by relating the~point $x(0)$ and the~terminal point of tracking $x(t_\mathrm{switch})$ to a reachability certificate. While this approach still consists of system simulations to evaluate reachability, these are not of full length since they do not require to reach $G$ as in~\cite{LQRtrees2}. Thus, they are far more efficient since $t_\mathrm{switch} \ll T$ in most cases as we demonstrate in our computational experiments.

We denote the tracking controller that we use to follow a demonstration $(\tilde{x},\tilde{u})$ by $u^{(\tilde{x},\tilde{u})}_{\mathrm{track}}$ which will refer as \emph{target demonstrations}. This tracking controller is generally a time-varying feedback control law $u_{\mathrm{track}}^{(\tilde{x},\tilde{u})} \colon \mathbb{R}^n \times [0,T] \mapsto \mathbb{R}^m$ and we assume that it is continuous function and that the corresponding solution of ODE~(1) with this tracking exists. Notably, we choose LQR tracking $u^{(\tilde{x},\tilde{u})}_{\mathrm{LQR}}$ in the implementation described in the next sections. 

When switching demonstrations, the~question is, which~new target demonstration to~choose. We will assume that~some predefined rule $\phi$ of the~following form was chosen.
 
\begin{Definition}
\label{assignmentrule}
An \emph{assignment rule} $\phi$ is a~function that~for any set of demonstrations $\mathcal{T}$ and point $x\in \mathbb{R}^n$ selects an~element from $\mathcal{T}$. 
\end{Definition}

Now we can define a~control law that~switches the~target demonstration to the~one selected by an~assignment rule in certain time intervals.

\begin{Definition}
\label{switchingcontrol}
Assume a~set of demonstrations $\mathcal{T},$ and for each demonstration $(\tilde{x},\tilde{u})\in\mathcal{T}$ a corresponding tracking controller $u_{\mathrm{track}}^{(\tilde{x},\tilde{u})}$. Assume an~assignment rule $\phi$. A~\emph{switching control law based on $\mathcal{T}$, the corresponding family of tracking controllers $u_{\mathrm{track}}$ and the assignment rule $\phi$} is a~function
$$\USwitch{\mathcal{T}}{\mathcal{U}}\colon \mathbb{R}^n\times [t_0, \infty)\rightarrow\mathbb{R}^m$$
such that~for every $x_0\in\mathbb{R}^n$ there~are  $t_1, t_2, \ldots$ (that~we call \emph{switching times}) and corresponding target demonstrations $(\tilde{x}_1,\tilde{u}_1), (\tilde{x}_2, \tilde{u}_2), \dots \in\mathcal{T}$  such that for every $j \geq 0$, $0 < t_{j+1}-t_j\leq T$, and for $t=t_0$, as well as for all $t\in\left[t_j,t_{j+1}\right)$, 
 \[
   \USwitch{\mathcal{T}}{\mathcal{U}}(x,t) = u_{\mathrm{track}}^{\phi(\mathcal{T},\Sol{x_0}{\USwitch{\mathcal{T}}{\mathcal{U}}}(t_j))}(x,t-t_j),\]
with $t_0=0$.

\end{Definition}

\begin{figure}
\centering
\includegraphics[width=0.85\linewidth]{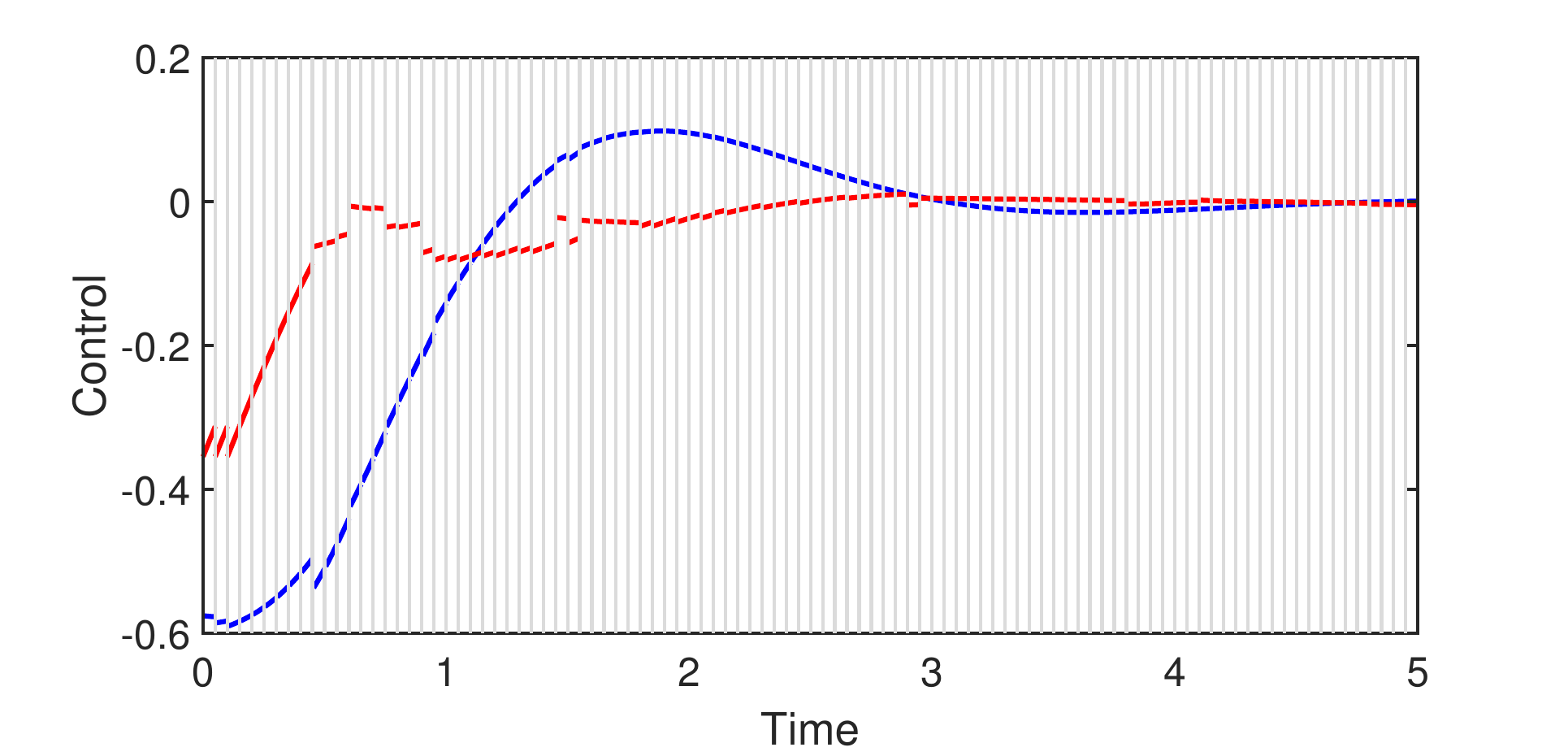}
\includegraphics[width=0.85\linewidth]{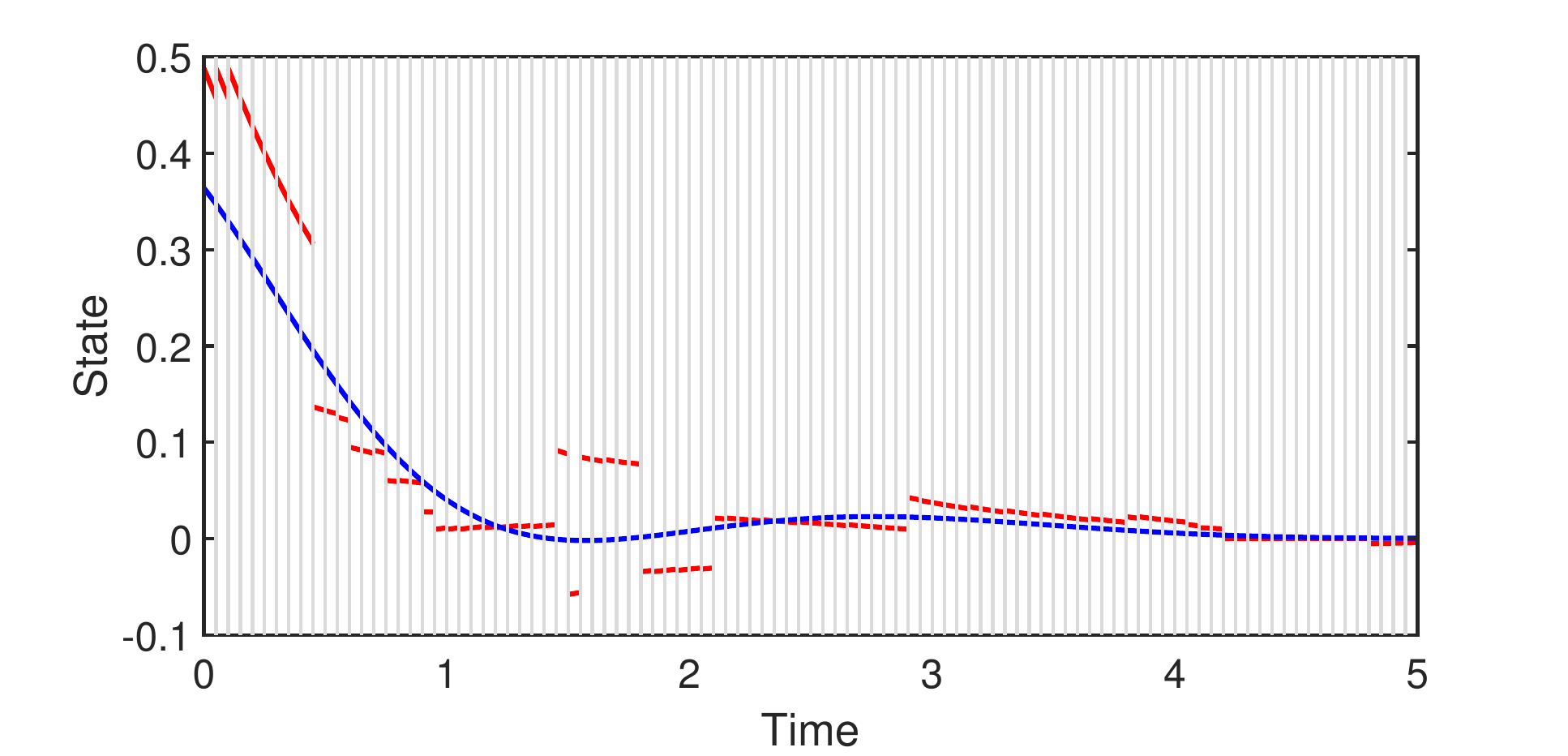}
\caption{Switching control input and the resulting switching trajectory (blue) and the corresponding parts of target demonstrations (red) with time intervals between switches (gray). The simulation of the quadcopter example (see description in Section 6.1., only one state and one control input) with the LQR switching controller produced by our implementation of Algorithm 2.}
\label{switching_traj}
\end{figure}

Note that this definition is recursive: It chooses the target demonstration used in the time interval $[t_j,t_{j+1})$ based on the state $\Sol{x_0}{\USwitch{\mathcal{T}}{\mathcal{U}}}(t_j)$, which is the state reached at time $t_j$ when using the switching controller up to this time. It uses the assignment rule $\phi$ to arrive at the target demonstration $\phi(\mathcal{T}, \Sol{x_0}{\USwitch{\mathcal{T}}{\mathcal{U}}}(t_j))$ corresponding to this state. This results in a discontinuous switch of control input at time $t_j$. An example of an LQR based switching controller can be seen in Figure~\ref{switching_traj}.

Now observe that~unlike an individual tracking, such switching control laws are generally \emph{not} unique due to the~non-unique switching times. However, we will narrow our interest to switching controllers that have switching times given by \emph{predefined} conditions, which when met, a next switch occurs.

To obtain these,  we will work with a continuous functions $L\colon \mathbb{R}^n \mapsto \mathbb{R}$ that will serve as a reachability certificate. The intuition is that this function will satisfy a Lyapunov-like decrease condition on demonstrations, which will then allow us to check reachability of the constructed tracking controller \emph{relative} to the behavior of the demonstration it follows between two switches.

The decrease condition on demonstrations is the following:

\begin{Definition}
\label{compatibility_def}
Assume a set of demonstrations $\mathcal{T}$ and a finite sequence $\theta = (\theta_1, \theta_2, \ldots, \theta_N )$ such that $0 < \theta_1 < \cdots < \theta_N = T$. A~function $L:\mathbb{R}^n\rightarrow\mathbb{R}$ is \emph{compatible} with~$\mathcal{T}$ at time instants $\theta$
iff there is $\delta<0$ such that for all $(\tilde{x},\tilde{u})\in\mathcal{T}$ and for all $j = 1, \ldots, N$
$$L(\tilde{x}(\theta_{j})) - L(\tilde{x}(\theta_{j-1})) \leq \delta .$$ 
Here $\theta_0=0$.
\end{Definition}

In this definition, we assume that demonstrations from $\mathcal{T}$ all are compatible with $L$ at the same time instants $\theta$ to keep the presentation simple. However, time instants $\theta$ can be chosen individually for each demonstration which is advantageous, since it eases the construction of such a function as we will discuss in Section \ref{sec:algorithm}.

We want to~use the function $L$ to~ensure reachability of the~resulting switching control law. For this, we will use a~criterion similar to a finite-time Lyapunov functions~\cite{Aeyels:98} that allows us avoid having to commit to an absolute minimal value for the decrease of the function $L$. Instead, we will ensure sufficient
decrease of $L$ relatively to the followed target demonstration based on a constant $1>\gamma>0$:
\begin{Definition}
\label{switchingcontrolL} 
A~switching control law $\USwitch{\mathcal{T}}{\mathcal{U}}$ is \emph{sufficiently decreasing} on $I$ wrt. a function $L: \mathbb{R}^n \rightarrow \mathbb{R}$, time instants $\theta$, and a goal set $G$ iff for every $x_0  \in I\setminus G$, a switching trajectory $x=\Sol{x_0}{\USwitch{\mathcal{T}}{\mathcal{U}}}$ with corresponding switching times $t_1,t_2,\dots$ and target demonstrations $(\tilde{x}_1,\tilde{u}_1), (\tilde{x}_2,\tilde{u}_2), \dots$  meets for every $j\geq 0$
\begin{itemize}
\item either $x(t_{j+1})\in G$,
\item or $x(t_{j+1})\in I$ and $L(x(t_{j+1}))  - L(x(t_j))\leq \gamma[L(\tilde{x}_j(t_{j+1}-t_j)) - L(\tilde{x}_j(0))]$ and $t_{j+1}-t_j \in \theta$
\end{itemize}
Here $t_0=0$.
\end{Definition}

The implementation of such a switching~control law consists of checking the switching conditions and switching demonstrations if these conditions are met, see Figure~\ref{switching_traj}. Since this switching~control law decreases $L$ at least by $\gamma\delta$ over one trajectory switch, the~system must visit $G$ after finitely many trajectory switches, otherwise we get a contradiction with the fact that $L$ must be bounded from below over $I\setminus G$. Hence we now have a suitable definition of the notion of reachability certificate.

\begin{Definition}
A continuous function $L: \mathbb{R}^n \rightarrow \mathbb{R}$ is a \emph{reachability certificate} for a switching control law $\USwitch{\mathcal{T}}{\mathcal{U}}$ based on a set of demonstrations $\mathcal{T}$ and a goal set $G$ iff 
\begin{itemize}
\item  it is compatible with $\mathcal{T}$, and
\item $\USwitch{\mathcal{T}}{\mathcal{U}}$ is sufficiently decreasing on $I$ wrt. $L$, $\theta$, $G$
\end{itemize}
\end{Definition}

Note that formal verification of this property would require solving the given ODE precisely. This is indeed possible~\cite{Nedialkov:99}, and the given definition simplifies the task due to its restriction to solutions between switches. As discussed, the certificate proves reachability.

\begin{Proposition}
\label{prop:cert}
The existence of a reachability certificate for a switching control law ensures that every system trajectory of the resulting closed loop system visits the goal set after finitely many trajectory switches.
\end{Proposition}

\subsection{Controller Synthesis Algorithm}
\label{sec:algorithm}

It is time to~instantiate the~general algorithm from Section~\ref{sec:generic} to the~switching control law from Definition~\ref{switchingcontrol}. The~result is  Algorithm~\ref{alg:unknown}. The~algorithm concretizes the~construction of a~generic control law  using a family of tracking controllers $u_{\mathrm{track}}$. 

\begin{algorithm}
  \caption{Controller and Compatible Certificate Synthesis}
  \label{alg:unknown}
{\small	
\begin{description}
\item[Input: ] ~
  \begin{itemize}
  \item a~control system of the~form stated in Section~\ref{sec:problem} with a~compact initial set $I$ and a~compact goal set $G$
  \item a~demonstrator generating demonstrations on some open superset $W$ of $I$
	\item a~set of reachability certificate candidates $\mathcal{L}$

  \end{itemize} 
\item[Output:] ~
  \begin{itemize}
  \item a~control law that steers the system from $I$ into $G$ and optionally satisfies some additional performance requirements, and 
  \item a reachability certificate $L$
  \end{itemize}
\end{description}

\begin{enumerate}
\item Generate an initial set of demonstrations  $\mathcal{T}$.
\item Choose a sequence  $M$ of states from  $W$ whose set of elements is a dense subset of  $W.$

\item Iterate:
  \begin{enumerate}
  \item\label{ref:synth_cert} Learn a~reachability certificate candidate $L \in \mathcal{L}$ compatible with $\mathcal{T}.$
  
 \item\label{step:decrease_check} Let $x_0$ be the next sample from the sequence $M$. Let $(\tilde{x}, \tilde{u})=  \phi(\mathcal{T}, x_0)$, $x=\Sol{x_0}{u_{\mathrm{track}}^{(\tilde{x}, \tilde{u})}}$. Check if there~is a~$t\in [0,T]$ s.t.
   \begin{itemize}
   \item either $t\in\theta$ and $x(t)\in I$ and $L(x(t))  - L(x(0)) \leq \gamma\left[L(\tilde{x}(t)) - L(\tilde{x}(0))\right]$
   \item or $x(t)\in G.$
   \end{itemize}
If this condition does not hold, generate a~new demonstration starting from $x_0$ and add it to $\mathcal{T}$.

\end{enumerate}
\item return a switching control law based on $\mathcal{T}$ with a reachability certificate $L$
\end{enumerate}
}
\end{algorithm}

The algorithm depends on the following parameters:
\begin{itemize}
\item a~tracking method that assigns to every demonstration $(\tilde{x},\tilde{u})$ a tracking controller $u_{\mathrm{track}}^{(\tilde{x},\tilde{u})}$
\item an assignment rule $\phi$ 
\item a real number $1>\gamma >0$ and time instants $\theta$ 
\end{itemize}

Let us explore the inner workings of Algorithm~\ref{alg:unknown} more closely. Algorithm~\ref{alg:unknown} tests whether the~constructed switching control law is sufficiently decreasing which, due to Proposition~\ref{prop:cert}, ensures reachability. A reachability certificate candidate $L$ is chosen to be compatible with demonstrations $\mathcal{T}$ that are generated as the algorithm progresses.

\begin{figure}
\center
\includegraphics[width=0.6\linewidth]{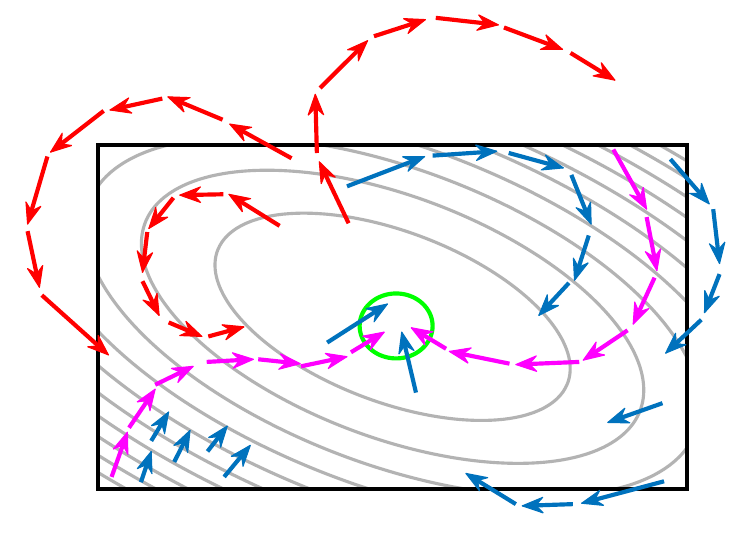}
\caption{System simulations in Algorithm~\ref{alg:unknown}: goal set (green ellipse), demonstrations (magenta), compatible reachability certificate candidate (contour plot) on the set of initial states $I$ (rectangle). Simulations that meet the sufficient decrease condition are blue, others are red.}
\label{system_sim}
\end{figure}

The algorithm adds a new demonstration whenever it detects via a system simulation (see Figure~\ref{system_sim}) a counterexamples, a state for which the resulting switching controller does not meet Proposition~\ref{prop:cert}. While the algorithm could decide whether to add a new demonstration purely by full simulations as in~\cite{LQRtrees2}, thanks to the learned reachability certificate candidate, the algorithm uses the sufficient decrease condition. In practice, this makes system simulations often significantly shorter, and thus it makes the overall run of the algorithm noticeably faster, as we will demonstrate in our numerical experiments. 

In addition, notice that the algorithm does test simulations from a larger set  $W \supset I\setminus G$. This is caused by the fact that algorithm checks the decrease condition via sampling and thus it requires an open investigated set. Otherwise, there could be some not removable counterexamples namely on the boundary of the investigated set. This also implies that the demonstrator must provide demonstrations from $W$ as well.

Finally, notice that Algorithm 2 as stated here does not state when to terminate the loop in step (3). We will show in the next section that the loop can be terminated eventually since reachability of the resulting control law on the whole set $I$ will be met. This improves earlier results~\cite{LQRtrees1,LQRtrees2} that only describe algorithm behavior for the number of iterations tending to infinity. See Section 5.4 on how to ensure termination of the loop in practice.

\subsection{Learning reachability certificate}
For our further considerations, we will assume that the set of reachability certificate candidates $\mathcal{L}$ is given in
parametrized form, that is, as a set $\mathcal{L} = \{L_p(x) \mid p\in P\}$, where  $L_p(x)$ is continuous in both $p$ and $x$ and the set of values of the parameters $P$ is compact. In such a case, learning (step 3a) consists of finding values of $p \in P$ and $\delta <0$ that meet a finite system of inequalities, $L_p(\tilde{x}(\theta_{j})) - L_p(\tilde{x}(\theta_{j-1})) \leq \delta$ for all $j = 1, \ldots, N$ and all $(\tilde{x},\tilde{u}) \in \mathcal{T}$.

\subsection{Algorithm Properties}
\label{sec:properties}

In this subsection, we will establish properties of Algorithm~\ref{alg:unknown} in terms of reachability and asymptotic optimality of the resulting switching control law. First, we show that, in finitely many iterations,  Algorithm~\ref{alg:unknown} creates a feedback control law that reaches the goal set. 

We start with repeating our standing assumptions. We assume smooth system dynamics~\eqref{system}, compact sets of initial states and goal states and a demonstrator, that provides demonstrations in terms of Definition 1. We need to add some more assumptions. We start with the set of reachability certificate candidates.
\begin{enumerate}
\item[(\,I\,)] We assume that the set of candidates $\mathcal{L}$ includes a candidate that is compatible with all demonstrations generated by the demonstrator as the Algorithm 2 progresses. 
\end{enumerate}
Assumption (I) is not that strict in practice since the time instants $\theta$ can be chosen individually for each demonstration and thus these instants can be placed sparsely for particular demonstrations if needed. However, to see the benefit of shortening the simulations, more expressive functions are required. These can follow shapes of demonstrations more closely and thus can be compatible for more densely placed $\theta$, which in turn leads to shorter system simulations in the falsification loop of the algorithm. In the implementation, we use a set of polynomials up to a given degree which proved to be sufficient to observe these benefits. 

Next, we add a technical, albeit crucial, assumption that ensures that the decrease condition is met while tracking any given demonstration closely enough. 
\begin{enumerate}
\item[(\,II\,)] We assume that there are is an $\varepsilon > 0$ such that any demonstration $(\tilde{x},\tilde{u})$ includes a point $\tilde{x}(t)$ for some $t\in\theta$ such that an $\varepsilon$-neighbourhood with the center in $\tilde{x}(t)$ lies in either $I$ or $G$.
\end{enumerate}

\begin{figure}
\center
\includegraphics[width=0.5\linewidth]{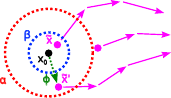}
\caption{Assumption (\,III\,): Assignment rule $\phi$ must always assign a sufficiently close demonstration (i.e. from a neighborhood $\alpha$) if at least one demonstration exists in a neighborhood $\beta$. }
\label{arule}
\end{figure}

Further, we need to guarantee that the assignment rule~$\phi$ always assigns sufficiently near demonstrations, see Figure~\ref{arule}.

\begin{Definition}
\label{dom}
We will say that an~assignment rule is \emph{dominated by the~Euclidean norm} iff for every $\alpha>0$ there~is a~$\beta>0$ such that~for every finite set of demonstrations $\mathcal{T}$ and every $x_0\in \mathbb{R}^n$, if there is a $(\tilde{x}, \tilde{u})\in \mathcal{T}$ with $\norm{x_0-\tilde{x}(0)}<\beta$, then there exists a~$(\tilde{x}', \tilde{u}')\in \mathcal{T}$ with $\phi(\mathcal{T}, x_0)= (\tilde{x}', \tilde{u}')$, and $\norm{x_0-\tilde{x}'(0)}<\alpha$.
\end{Definition}

\begin{enumerate}
\item[(\,III\,)] We assume that the~assignment rule $\phi$ used in Algorithm~\ref{alg:unknown} is dominated by the~Euclidean norm.
\end{enumerate}

Algorithm~\ref{alg:unknown} detects counterexamples---states that do not meet the decrease condition---via sampling using the sequence of states $M$. Hence, we need to make certain that the presence of a counterexample implies the existence of a whole open ball of counterexamples in its vicinity. Due to continuity of ODEs solution wrt. initial conditions, there is a open ball of counterexamples around any given a counterexample provided that all states in this neighbourhood are assigned to the same demonstration. However, this cannot always be the case since any assignment rule is necessarily discontinuous for a finite set of demonstrations. Consequently, we need to further restrict the assignment rule $\phi$.

\begin{enumerate}
\item[(\,IV\,)] We assume that for any finite set of demonstrations $\mathcal{T}$, and for any state $x\in \mathbb{R}^n$, and any $\varepsilon >0$, there is an open ball that is a subset of an $\varepsilon$-neighbourhood of $x$ and includes only states that are assigned via $\phi$ to the same trajectory as $x$.
\end{enumerate}

Notice that an assignment rule based on the Euclidean distance meets Assumption (\,IV\,) since such an assignment rule partitions $\mathbb{R}^n$ into Voronoi polygons. 

The last assumption assures that the tracking controller  $u_{\mathrm{track}}^{(\tilde{x}, \tilde{u})}$ used in Algorithm 2 follows the target demonstrations sufficiently well.

\begin{enumerate}
\item[(\,V\,)] For any $\varepsilon > 0$ there is a $\delta >0$ such that for any $x_0\in\mathbb{R}^n$ and for any demonstration $(\tilde{x},\tilde{u})$ with $\norm{x_0 - \tilde{x}(0)}< \delta$ and for all $t\in[0, T]$,
$$\norm{\Sol{x_0}{u_{\mathrm{track}}^{(\tilde{x}, \tilde{u})}}(t) - \tilde{x}(t)} < \varepsilon.$$
\end{enumerate}

We will prove that Assumption (\,V\,) holds for LQR tracking in the next subsection. With all assumptions stated, we can state the main results of this section. Assumptions (\,I\,)--(\,V\,) assure that there is a neighbourhood around each initial point of any added demonstration, where the sufficient decrease condition is met. Thus, Algorithm~\ref{alg:unknown} eventually covers the whole set $W$ with demonstrations in such a way that the sufficient decrease condition is met everywhere in $W$, which implies reachability due to Proposition~\ref{prop:cert}. See proof of Proposition \ref{prop:alg_reach} in the~appendix for more details.

\begin{Proposition}
\label{prop:alg_reach}
Assume that Algorithm~\ref{alg:unknown} uses the tracking controllers $u_{\mathrm{track}}$ and let Assumptions (\,I\,)--(\,V\,) hold. Then, in finitely many iterations, Algorithm~\ref{alg:unknown} provides a~control law that steers the system from $W$ into $G$. 
\end{Proposition}

We can also establish asymptotic relative optimality of the switching control law, provided that the demonstrator is continuous and consistent.

\begin{Definition}
A demonstrator is
\begin{itemize}
\item \emph{continuous} iff the provided control input $u$ is jointly continuous in both the~initial state $x_0\in W$ and time $t\in [0, T]$, and
\item \emph{consistent} iff for all demonstrations $(x_1,u_1)$ and  $(x_2,u_2)$ provided by the demonstrator and all $t_1,t_2\in [0,T]$, if $x_1(t_1)=x_2(t_2)$, then $u_1(t_1)=u_2(t_2)$.
\end{itemize}
\end{Definition}
Continuity of the demonstrator assures that demonstrations that start from near states are similar. Consistency of the demonstrator results in unique cost associated with the demonstrator for each state of the system. A consistent demonstrator also allows for more efficient generation of demonstrations, since each system trajectory with inputs $u$ generated by the demonstrator can be interpreted as an infinite number of demonstrations, each one starting from a different state of the generated trajectory. In addition, consistency allows us to use a switching controller without loosing relative optimality.  

\begin{Proposition}
\label{prop:cost}
Assume a finite set of demonstrations~$\mathcal{T}$ provided by a continuous and consistent demonstrator and an~assignment rule that is dominated by the~Euclidean norm. Let
$\USwitch{\mathcal{T}}{\mathcal{U}}$ be a switching controller based on $u_{\mathrm{track}}$ that meets Assumption (\,V\,), and let $(\tilde{x},\tilde{u})$ be a demonstration. Then for all $t\in [0, T]$, 
$$ \lim_{x_0 \rightarrow \tilde{x}(0)}\left|\Cost{V_t}{x_0}{\USwitch{\mathcal{T}}{\mathcal{U}}} - \Cost{V_t}{\tilde{x}(0)}{\tilde{u}}\right| = 0.$$
\end{Proposition}
The following corollary extends this performance guarantee to the whole set $W$. 
\begin{Corollary}
\label{cor:optimality}
Assume a continuous and consistent demonstrator, an~assignment rule that is dominated by the~Euclidean norm and a tracking controller $u_{\mathrm{track}}$ that meets Assumption (\,V\,). Let $x_0^1,x_0^2,\dots$ be a~sequence of states in  $W$ whose set of elements is a dense subset of  $W$. Let $(\tilde{x}_1,\tilde{u}_1)$, $(\tilde{x}_2,\tilde{u}_2)$, $\dots$ be the~sequence of corresponding demonstrations generated by the~demonstrator for $\tilde{x}_1(0)= x_0^1, \tilde{x}_2(0)= x_0^2, \dots$. For $k= 1, 2 \dots$, let $\mathcal{T}_k= \{ (\tilde{x}_i,\tilde{u}_i) \mid i\in \{ 1,\dots,k\}\}$, and let $\USwitch{\mathcal{T}_k }{U}$ be a corresponding switching control law for $u_{\mathrm{track}}$. Then
\[ \lim_{k\rightarrow \infty} \left|\Cost{V_t}{x_0}{\USwitch{\mathcal{T}_k }{U}} - \Cost{V_t}{x_0}{u}\right| \rightarrow 0 \]
for all $x_0 \in W$ and for all $t\in[0, T].$
\end{Corollary}
In particular, the~limit converges to an optimal controller, if the~demonstrator provides optimal demonstrations. Consequently, if we sample the~state space thoroughly enough, the~cost of the~control law will approach the~cost of the~demonstrator, and hence we can ensure the~additional performance requirements by stating the~termination condition of the~algorithm via performance comparison with the~demonstrator.

While the introduced properties of the algorithm ensure reachability in finitely many iterations, and relative optimality in the limit, they do not state an explicit criterion for terminating the loop. In industry, safety-critical systems are often verified by systematic testing. This can be reflected in the~termination condition for the~main loop by simply considering each loop iteration that~does not produce a~counterexample as a~successful test. Based on this, in a similar way as related work~\cite{LQRtrees2}, one could terminate the~loop after the~number of loop iterations that~produced a~successful test without any intermediate reappearance of a~counterexample exceeds a~certain threshold. One could also require certain coverage~\cite{Dang:09} of the~set $I$ with successful tests. And for even stricter safety criteria, one can do formal verification of the~resulting controller.

\subsection{Algorithm 2 with LQR tracking}

For our further discussion, we will reduce our attention to a single type of tracking, specifically the linear-quadratic regulator (LQR), for which~we refer the reader to the~literature~\cite{LQRtrees1,LQRtrees2}. We will denote by~$\LQRTrack{(\tilde{x},\tilde{u})}$ an LQR tracking controller of a demonstration $(\tilde{x},\tilde{u})$. This LQR controller is characterized by a solution of the differential Riccati equation. Further, we will assume that all cost matrices used in the construction of LQR tracking are positive definite and fixed for all demonstrations.

First, we prove that LQR tracking meets Assumption (\,V\,). The proof follows the structure of the proof of Lemma~4 in Tobenkin~et.~al.~\cite{SOS_funnels} and the~reader can find in the~appendix. 
\begin{Lemma}
\label{keylemma}
For any $\varepsilon > 0$ there is a $\delta >0$ such that for any $x_0\in\mathbb{R}^n$ and for any demonstration $(\tilde{x},\tilde{u})$ with $\norm{x_0 - \tilde{x}(0)}< \delta$ and for all $t\in[0, T]$,
$$\norm{\Sol{x_0}{\LQRTrack{(\tilde{x},\tilde{u})}}(t) - \tilde{x}(t)} < \varepsilon.$$
\end{Lemma}
Hence, Algorithm 2 with a tracking controller $\LQRTrack{(\tilde{x},\tilde{u})}$ provides a control law that reaches the goal set after finitely many iterations and is asymptotically relatively optimal provided that the demonstrator is continuous and consistent. Moreover, after dropping Assumption (\,I\,), this result also holds for the case without any use of certificates which corresponds to earlier methods~\cite{LQRtrees1,LQRtrees2} that were published without such guarantees. Reist et. al.~\cite{LQRtrees2} write ``However, while the analysis shows that, in the long run, the algorithm tends to improve the generated policies, there are no guarantees for finite iterations.''

The LQR tracking controllers have another benefit in providing an assignment rule that respects the dynamics of the system where as the assignment rule based on the Euclidean distance does not. Namely, we set for a finite set of demonstrations $\mathcal{T} = \{(\tilde{x}_1, \tilde{u}_1), \ldots, (\tilde{x}_N, \tilde{u}_N)\}$
\begin{equation}
\label{LQRarule}
\phi(x,\mathcal{T}) = \underset{{i = 1,\ldots, N}}{\mathrm{arg\,min}} \left\{(x-\tilde{x}_i(0))^TS_i(0)(x-\tilde{x}_i(0))\right\}
\end{equation}
where $S_i$ is a solution to the appropriate differential Riccati equation. Note that rule $\eqref{LQRarule}$ is actually not a properly defined assignment rule, since it allows ties. These are situations in which a state $x$ could be assigned to multiple distinct demonstrations, since $|\phi(x,\mathcal{T})| \neq 1$. These ties must be resolved by assigning each state to a single demonstration from $\phi(x,\mathcal{T})$. The resolution of these ties is closely connected to assumption (\,IV\,).
\begin{Lemma}
Assignment rule~\eqref{LQRarule} with any resolution of ties is dominated by the Euclidean norm, and meets assumption (\,IV\,) for some appropriate resolution of ties.
\end{Lemma}

\section{Implementation}
\label{sec:implementation}
\label{FID}

In this section, we describe our implementation of Algorithm~\ref{alg:unknown} which we will refer as LQR switching algorithm. The~whole implementation was done in MATLAB 2020b. We used CasADi~v3.5.5~\cite{Andersson2018} with the optimization solver Ipopt~\cite{ipopt} with their respective default settings for computation of demonstrations. System simulations were numerically approximated using the RK4 integrator with the same time step used in the computation of demonstrations.

\begin{description}[style=unboxed,leftmargin=0cm]
\setlength\itemsep{1em}
\item[Demonstrator] The demonstrator is based on a~direct optimal control solver~\cite{Bet:10} that~solves the~nonlinear programming obtained from the~optimal control problem by discretization, namely the~Hermite-Simpson collocation method \cite{Har:87}. The~initial solutions for the~demonstrator were set as constant zero for both states and control.

\item[Time instants $\theta$] To extract as much information from the demonstrator as possible, we compute demonstrations consisting of $2N-1$ discrete samples with an equidistant time step $h$. These time instants also serve as our choice of the time instants $\theta$. From these original demonstrations, we generate $N$ demonstrations of length $N$, the $i$-th demonstration starting from the $i$-th sample. This means that with each added demonstration, actually multiple demonstrations are added, ensuring consistency according to Definition~\ref{demonstrator_def}.
\item[Reachability certificate] We use a linearly parametrized set of certificate candidates: a set of polynomials. We choose the maximum degree for each problem in such a way that compatible reachability certificate candidate always existed during the computational experiments. Since candidates are parametrized linearly, we can pick a candidate as any solution of the system of linear inequalities. In order to pick a reasonable one, we select the Chebyshev center~\cite{Rav:19} of the polytope described by this system of linear inequalities. This selection strategy requires merely a solution of a linear program~\cite{Rav:19}.

In addition, we use for learning merely samples of demonstrations that lie in $W$, since the decrease condition requires that the system lies in $W$ anyway. If some values on a demonstration outside of $W$ are required, we merely estimate them using the closest known values on the same demonstration.
\item[Assignment rule] We use assignment rule~\eqref{LQRarule} with a slight modification: we first narrow the~search of the~target demonstration to the 100 closest demonstrations according to the Euclidean distance. This significantly speeds up the assignment for high dimensional examples with large number of generated demonstrations.
\item[Termination of the loop] We use use the~following termination criterion for the main loop:  Finish the~loop, accepting the~current strategy, if the~number of simulations without finding a~counterexample exceeds a~certain threshold $N_\mathrm{sim}$.
\end{description}

\begin{figure}
\center
\includegraphics[width=0.8\linewidth]{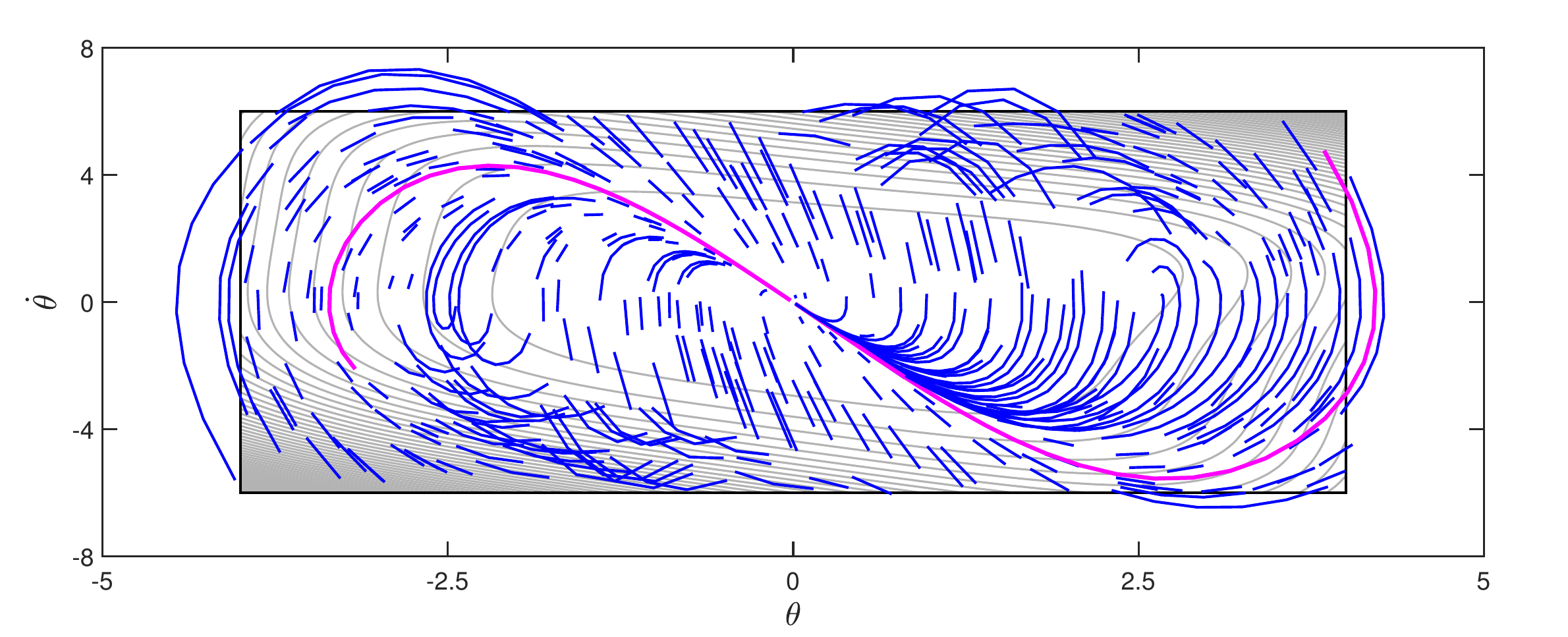}
\caption{Several iterations of LQR switching algorithm for the pendulum example: demonstrations (magenta), reachability certificate candidate (contour plot) on the set of initial states $I$ (inner rectangle) compatible with these demonstrations. System simulations in Algorithm 2 (blue) are terminated, when the sufficient decrease condition is met.}
\label{alg}
\end{figure}

\section{Computational Experiments}

\begin{table}
\center
\begin{tabular}{l  || l l || c  c c }
\hline
Ex.  &  \multicolumn{1}{l}{$T$} & \multicolumn{1}{l||}{$h$} & deg  & \multicolumn{1}{l}{$N_\mathrm{sim}$} & $Q,R$  \\
\hline	
Pnd. & 10 & 0.05 & 4 & $2\cdot10^5$ & $I_2,I_1$\\
Pndc. & 10 & 0.05 & 2  & $4\cdot10^5$ & $I_4,I_1$\\
RTAC & 150 & 0.05 & 4 & $4\cdot10^5$ & $I_4,I_1$\\
Acr. & 10 & 0.01 & 4 & $4\cdot10^5$ & $100I_4,I_1$\\
Fan & 40 & 0.05 & 2 & $6\cdot10^5$  & $I_6,I_2$\\
Quad. & 20 & 0.05  & 2 & $12\cdot10^5$ & $I_{12},I_4$\\
\hline
\multicolumn{6}{l}{}\\
\end{tabular}
\caption{Parameters of demonstrations and the LQR switching algorithm: length of demonstrations $T$, time step $h$, degree of the polynomial used for reachability certificate candidates, number of successful subsequent simulations to accept $N_\mathrm{sim}$, LQR cost matrices $Q$ and $R$}
\label{tab:par}
\end{table}

\begin{table*}
\center
\begin{tabular}{l  r r r  r}
Ex. &  $N_\mathrm{dem}$ & $l_\mathrm{sim}$ & $t_\mathrm{sim}$ & $t_{\mathrm{total}}$\\
\hline	
Pnd. & 2 & 0.15  & 0.07  & 0.17 \\
Pndc. & 14 & 1.65 & 2.79 &  3.50 \\
RTAC & 4 & 2.75 & 5.04 & 10.92 \\
Acr.& 3 & 1.85 & 5.80 & 7.04 \\
Fan & 22 & 1.80 & 3.84 & 8.57\\
Quad. & 98 & 1.64 & 28.01 & 49.41\\
\hline
Pnd. & 3  & 1.91  & 0.53   & 0.63 \\
Pndc. & 14  & 5.36   & 8.13  & 8.83 \\
RTAC  & 11  & 73.18   & 213.87   & 227.60 \\
Acr. & 3  & 3.88  & 11.64  & 12.85 \\
Fan & 24 & 24.15  & 43.52  & 48.23 \\
Quad. & 111  & 7.53  & 154.81  & 178.63\\
\hline
Pnd. &  2 & 1.91 & 0.57 & 0.67  \\
Pndc. & 6 & 5.34 & 7.12 & 7.54\\
RTAC &  TO & TO & TO & TO\\
Acr.& 1 & 3.90 & 14.50 & 15.12\\
Fan & 21 & 24.31 & 304.04 & 309.93\\
Quad. & $^*$73 & $^*$7.53 & $^*$562.69& $^*$590.25\\
\hline
\multicolumn{5}{l}{}\\
\end{tabular}
\caption{LQR-tree algorithm with a reachability certificate (top), and simulation-only algorithm~\cite{LQRtrees2} with assignment rule~\eqref{LQRarule} (middle) and assignment rule~\eqref{LQRarulemod} (bottom): number of demonstrations added $N_\mathrm{dem}$, average lengths of simulations $l_\mathrm{sim}$, CPU time in minutes for simulations $t_\mathrm{sim}$, total CPU time in minutes $t_{\mathrm{total}}$. A star denotes that some runs exceed 10 hours CPU time mark (TO). All values are averages using 5 runs of the algorithms with different seeds of state samples.}
\label{tab:time}
\end{table*}

\begin{table}
\center
\begin{tabular}{l  || *5{S[table-format=<4, table-space-text-post=\,*,table-align-text-post = false]}}
Ex.  & {Dem} & {LQRsw} & {LQRtr} & {LQReq}\\
\hline	
Pnd.& 22 & 24 & 24 & 43\\
Pndc.&92 & 98 & 98 & 98\\
RTAC &297& 401 & 430 & 636 \,*\\
Acr. &2516& 7656& 7669 & 8719\\
Fan &32 & 37& 37 & 72 \,* &\\
Quad. & 45 & 50 & 50 & 69&\\
\hline
\multicolumn{4}{l}{}\\
\end{tabular}
\caption{Mean cost on 1000 random samples: demonstrator, LQR switching controller (the result of Algorithm~\ref{alg:unknown}), LQR tracking controller (the result of algorithm~\cite{LQRtrees2}), LQR controller around equilibrium. A star denotes that the controller failed from some samples, which were discarded from the mean value. All values are averages using 5 runs of the algorithms.}
\label{tab:cost}
\end{table}

\label{sec:Examples}

In this section, we provide computational experiments documenting the~behavior of our algorithm and implementation on six typical benchmark problems. The problems range from dimension two to twelve and will concern the task of reaching a neighbourhood of an equilibrium. The cost functional is quadratic $V_t(x,u) = \int_0^t x(\tau)^Tx(\tau) + u(\tau)^Tu(\tau) \,\mathrm{d}\tau$.

In addition to demonstrating wide applicability of the control law synthesis algorithm bases on demonstrations that is supported by our theoretical analysis, we are mostly interested in examining the effect of a reachability certificate. The presence of this certificate should significantly speed up the simulation process. Thus, we compare our LQR switching algorithm with the most similar algorithm from the literature: simulation based LQR-trees~\cite{LQRtrees2}. This algorithm also employs demonstrations and LQR tracking, but neither uses a reachability certificate nor switching. Thus, it searches for counterexamples to reachability using full system simulations. 

Another slight difference is in the choice of the assignment rule~\eqref{LQRarule} based on estimation of funnels~\cite{LQRtrees2}. Let us denote $V_i(x) = (x-\tilde{x}_i(0))^TS_i(0)(x-\tilde{x}_i(0))$. Then the assignment rule in~\cite{LQRtrees2} is
\begin{equation}
\label{LQRarulemod}
\phi(x,\mathcal{T}) =\underset{{i = 1,\ldots, N}}{\mathrm{arg\,min}} \left\{V_i(x) \text{ s.t. } V_i(x) \leq d_i\right\},
\end{equation}
where the values $d_i$ are learned during the run of the algorithm: when the algorithm detects a counterexample via a full system simulation, instead of adding a new demonstration immediately, the algorithm  lowers the value of the corresponding $d_i$ such that the counterexample state is no longer assigned to the demonstration which led to failed reachability. Then the algorithm tries to assign a new demonstration via~\eqref{LQRarulemod} with the modified values of $d_i$s. The algorithm adds a new demonstration only when no current demonstration can be assigned to the sampled point. 

These modifications are still within the theory presented here and hence, the algorithm~\cite{LQRtrees2} still meets Proposition \ref{prop:alg_reach} and Corollary \ref{cor:optimality}, i.e. constructs a control law that achieves reachability in finitely many iterations and that is asymptotically relatively optimal for increasing number of demonstrations. For the sake of completeness, we will do our comparison between our algorithm and the full simulation algorithm~\cite{LQRtrees2} in two versions, with assignment rule~\eqref{LQRarulemod} and with assignment rule~\eqref{LQRarule} (both with a speed up using the Euclidean distance described in section~\ref{sec:implementation}). 

Lastly, to make these examples (except the acrobot example) a bit more challenging , we also required the synthesized controller to stay within a prespecified simulation region $S \supset I$. We also added this condition as a constraint on demonstrations to the demonstrator.

\subsection{Examples}
\begin{description}[style=unboxed,leftmargin=0cm]
\setlength\itemsep{1em}
\item[Inverted pendulum] We use the~model $\ddot{\theta} = \frac{g}{l}\sin\theta - \frac{b\dot{\theta}}{ml^2}  + \frac{u}{ml^2}$, where $x = \left[ \theta, \dot{\theta}\right]$ and $m = 1, l = 0.5, g = 9.81, b = 0.1$. We set the initial set $I = \left[-4, 4\right]\times\left[-6, 6\right]$ extended to $W = \left(-4.05, 4.05\right)\times\left(-6.05, 6.05\right)$, and set the goal set $G = \{x\in\mathbb{R}^2 \mid \norm{x} < 0.05\}$. Additionally, we set $S = \left[-5, 5\right]\times\left[-8,8\right]$ beyond which both learning algorithms declare simulations as unsuccessful. 

\item[Inverted pendulum on a~cart] We use the~model of the pendulum on cart~\cite{LQRtrees2} in which we set $m = 0.21, M = 0.815, g = 9.81$ a $l = 0.305$. We set the initial set (the order of variables is $[ x, \theta, \dot{x}, \dot{\theta}]$) $I = \left[-2.5, 2.5\right] \times \left[-2, 2\right] \times \left[-2.5, 2.5\right] \times \left[-2.5, 2.5\right]$ extended to $W = \left(-2.55, 2.55\right) \times \left(-2.05, 2.05\right) \times \left(-2.55, 2.55\right) \times \left(-2.55, 2.55\right)$, and set the goal set $G = \{x\in\mathbb{R}^4 \mid \norm{x} < 0.1\}$. Additionally, we set $S = \left[-8, 8\right] \times \left[-3, 3\right] \times \left[-12, 12\right] \times \left[-8, 8\right]$.

\item[RTAC (Rotational/translational actuator)] We use the~model \cite{RTAC} of dimension $4$. We set the initial set $I = \left[-5, 5\right]^4$ extended to $W = \left(-5.1, 5.1\right)^4$, and set the goal set $G = \{x\in\mathbb{R}^4 \mid \norm{x} < 0.05\}$. Additionally, we set $S = \left[-8.5, 8.5\right]^4$.

\item[Inverted double pendulum (acrobot)] We use the~same model as in \cite{acrobot} of dimension $4$. We set the initial set $I = \left[-0.17, 17\right]^4$ extended to $W = \left(-0.175, 175\right)^4$, and set the goal set $G = \{x\in\mathbb{R}^4 \mid \norm{x} < 0.0001\}$. In this particular example, we did not specify the set $S$.

\item[Caltech ducted fan] We use the~planar model of a~fan in hover mode~\cite{fan} of dimension $6$.  We set the initial set $I = \left[-1, 1\right]^6$ extended to $W = \left(-1.05, 1.05\right)^6$, and set the goal set $G = \{x\in\mathbb{R}^6 \mid \norm{x} < 0.01\}$. Additionally, we set $S = \left[-5, 5\right]^6$. 

\item[Quadcopter] We use the simplified~quadcopter model from~\cite[(2.61)--(2.66)]{quad_model}, where we set $m = 1$ and $g = 9.8$. We set the initial set $I = \left[-1, 1\right]^{12}$ extended to $W = \left(-1.05, 1.05\right)^{12}$ and set the goal set $G = \{x\in\mathbb{R}^{12} \mid \norm{x} < 0.01\}$. Additionally, we set $S = \left[-6, 6\right]^{12}$. 
\end{description}

\subsection{Results}

The problem-dependent parameters that we used are listed in Table~\ref{tab:par}.  We ran all algorithms for each example five times using five fixed sequences of samples $M$ generated via an uniform pseudorandom generator on the corresponding set $W$. The comparison of our algorithm based on LQR switching, and the full simulation based algorithms can be found in Table~\ref{tab:time}. All algorithms with a few exceptions were able to solve all benchmarks within our specification demonstrating their practical applicability supported theoretically by Proposition 2.

If we compare Algorithm~\ref{alg:unknown} with LQR tracking to both variants of the full simulation algorithm, our algorithm provides noticeable time savings in system simulations (between 50\% and 95\%). This is because Algorithm~\ref{alg:unknown} can declare a system simulation successful as soon as the decrease condition is met. Since the majority of computational time consists of system simulations, we can notice a significant time-savings in the overall runs of the algorithm as well. 

In addition, we observe a significant difference between full simulation algorithms. This indicates room for a simple improvement of the algorithm~\cite{LQRtrees2} by using our assignment rule~\eqref{LQRarule}. In the case of our benchmarks, the usage of the original assignment rule~\eqref{LQRarulemod} leads to a smaller number of added demonstration. However, the overall number of system simulations increased noticeably (except for the pendulum on a cart example) due to a larger number of detected counterexamples, so much so that we encountered a runtime larger than 10 hours for RTAC and quadcopter examples. Hence, it is a better strategy to simply keep adding new demonstrations whenever a counterexample is detected rather than modifying the assignment rule in order to assign a different existing demonstration, which is what~\cite{LQRtrees2} does.

\begin{figure}
\centering
\includegraphics[width=0.75\linewidth]{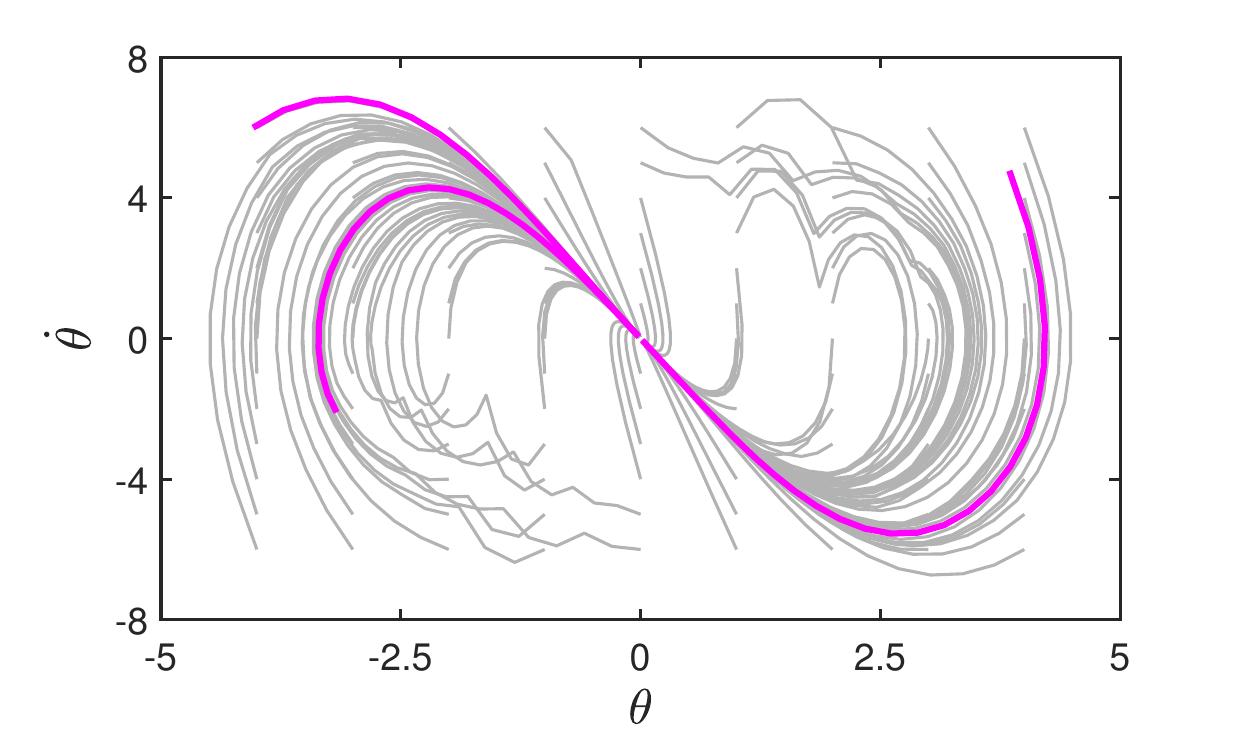}
\caption{Solution produced by LQR switching algorithm: demonstrations (magenta) and switching trajectories (gray).}
\label{pend_traj}
\end{figure}

We also evaluated the cost of the resulting controllers for 1000 random samples, see Table~\ref{tab:cost}. For comparison, we also include the cost of the demonstrator and the cost of an LQR controller computed around the equilibrium (with cost matrices from Table~\ref{tab:par}). The cost of LQR switching control law is quite similar to controller produced by the full simulation algorithm~\cite{LQRtrees2}. This illustrates that trajectory switching did not compromise the overall performance and thus, Algorithm~\ref{alg:unknown} provides a control law of similar performance to the full simulation algorithms but significantly faster. Notice also that the LQR control law around the equilibrium has a higher mean value cost in comparison to the both tracking based LQR controllers, which shows that using demonstrations significantly decreased the cost as supported by Corollary 1. Lastly, we also provide the resulting switching trajectories for the pendulum example in Figure~\ref{pend_traj}. 

\section{Conclusion}
\label{sec:conclusion}

We have proposed a~ offline framework for learning reachability certificates and synthesizing control laws that~inherits optimality with respect to a~chosen cost functional from the~given demonstrator. We have stated a~specific LQR variant of this framework and provided guarantees in terms of reachability in finitely many iterations and asymptotic relative optimality. We have also implemented the~algorithm and tested it on various case studies. We have made a~time comparison with a~full simulation approach without reachability certificates and seen significant time savings in the offline runtime of the algorithm (between 50\% and 95\%, which corresponds to reducing several hours of computation time to several minutes in the most extreme cases), while producing controllers of similar performance.

\bibliographystyle{plain}
\bibliography{ref} 
     
\appendix
\section{LQR tracking}
First, we give a brief overview of LQR tracking. Let $A(t) = A(\tilde{x}(t),\tilde{u}(t))$ and $B(t) = B(\tilde{x}(t),\tilde{u}(t))$ be the linearization of the system dynamics $F$ around the demonstration $(\tilde{x},\tilde{u})$. The LQR tracking is
\begin{equation}
\LQRTrack{(\tilde{x},\tilde{u})}(x,t) = -R^{-1}B(t)^TS(t)(x-\tilde{x}(t)) + \tilde{u}(t),
\end{equation} 
where $S(t)$ that meets the differential Riccati equation
\begin{equation}
\label{proof_riccati}
-\dot{S} = Q + SA(t) + A(t)^TS - SB(t)R^{-1}B(t)^TS
\end{equation}
with the terminal condition $S(T) = Q_0$, where $Q, R$ and $Q_0$ are all positive definite matrices. Since $V^{(\tilde{x},\tilde{u})}(x,t) = (x-\tilde{x}(t))^T S(t) (x-\tilde{x}(t))$ represents the optimal cost of tracking for the linearized dynamics, we can make the following observation which we use in our proofs. 

There are constants $c_1,c_2 >0$ such that for any $\LQRTrack{(\tilde{x},\tilde{u})}$ of any demonstration $(\tilde{x},\tilde{u})$,
\begin{equation}
\label{A5}
c_1I_n \leq S(t) \leq c_2I_n\text{ for all } t\in[0, T],
\end{equation}
where $I_n$ denotes the $n\times n$ identity matrix and the inequalities are in the sense of positive definiteness.

The inequality $S(t)\leq c_2I_n$ holds for some $c_2 > 0$, since the optimal cost $V^{(\tilde{x},\tilde{u})}(x,t)$ is bounded by the cost of evolutions using zero tracking inputs which is bounded, since $A(\tilde{x}(t),\tilde{u}(t))$ is bounded. As for the second inequality, matrix $S(t)$ is positive definite for all $t<T$ for any demonstration $(\tilde{x},\tilde{u})$, since $S(T) = Q_0$. Moreover, the eigenvalues of $S(t)$ cannot be arbitrarily small, since $Q$ and $R$ are positive definite and $A(\tilde{x}(t),\tilde{u}(t))$ and $B(\tilde{x}(t),\tilde{u}(t))$ are bounded. Thus $\Delta x(t)$ cannot become arbitrarily small arbitrarily fast using arbitrarily small control inputs, i.e., lead to arbitrarily small cost.

\section{Proof of Proposition 2} 
\begin{customProposition} 
Assume that Algorithm~\ref{alg:unknown} uses the tracking controllers $u_{\mathrm{track}}$ and let Assumptions (\,I\,)--(\,V\,) hold. Then, in finitely many iterations, Algorithm~\ref{alg:unknown} provides a~control law that steers the system from $W$ into $G$.  
\end{customProposition}
\begin{proof} We split the proof into three parts. In the first part, we provide the overall proof of Proposition 2 itself using two auxiliary observations shown in parts two and three.

\textbf{1.} The proof of this proposition is based on the following fact. Due to Assumptions (\,I\,)--(\,III\,) and  Assumption (\,V\,), there is $\delta>0$ such that for any $x\in W$, any finite set of demonstrations $\mathcal{T}$ and any reachability certificate candidate $L_p$ compatible with $\mathcal{T}$, all states from the $\delta$-neighborhood of $x$ meet decrease condition~\eqref{switchingcontrolL} provided that $\mathcal{T}$ contains a demonstration $(\tilde{x},\tilde{u})$ with  $\tilde{x}(0) = x.$ Let us for now assume that this $\delta>0$ indeed exists. We will provide the proof of this fact in the second part.

Now, assume a run of Algorithm~\ref{alg:unknown}. As the algorithm progresses, it adds demonstrations whenever the decrease condition is not met. However, there is a $\delta$-neighborhood around each initial point of any added demonstration, where the algorithm adds no further demonstration, since the decrease condition is met there. Consequently, only finitely many demonstrations can be added by the algorithm, because the investigated set $W$ is bounded. 

Let us consider the iteration, when the last demonstration is added. Let us assume that there are still counterexamples on $W$ to arrive at a contradiction from the assumption. As we will show in the third part of this proof, the existence of any counterexample to decrease condition~\eqref{switchingcontrolL} implies the existence of a whole open ball of counterexamples. However, this open ball consists samples from the rest of the sequence $M$ since it is still dense subset of $W$. Hence, the algorithm will add further demonstrations during the rest of the run, which is a contradiction. This completes the proof.

\textbf{2.} Here, we will show the existence of bound $\delta$ that we used in the first part. We combine several observations.

\begin{enumerate}
\item[(O1)] Candidates $\{L_p \mid p\in P\}$ are continuous in both $x$ and $p$ on a compact set $\bar{W} \times P$, where $\bar{W}$ denotes the closure of $W$. Thus, there is  $\delta'>0$ such that 
\begin{equation}
L_p(x_1) - L_p(x_2) \geq \gamma(L_p(x_1') - L_p(x_2'))
\end{equation}
for any $x_1,x_1',x_2,x_2'\in W$ and $p\in P$, for which $\norm{x_1-x_1'}<\delta'$ and $\norm{x_2-x_2'}<\delta'$.
\item[(O2)]  The assignment rule $\phi$ is dominated by the Euclidean norm.
\item[(O3)]  Assumption (\,V\,) holds.
\item[(O4)]  Assumption (\,II\,) holds.
\end{enumerate}
We just need to combine observations (O1)--(O4). There is a $\delta$-neighborhood of any initial state of any demonstration such that all states in this neighborhood are assigned to close demonstration due to (O2). These states evolve close to their assigned demonstrations due to (O3) and these evolutions reach $W$ or $G$ due to (O4). There, the decrease conditions are met due to (O1).
\end{proof}

\textbf{3.} Lastly, we show that the existence of any counterexample to decrease condition~\eqref{switchingcontrolL} implies the existence of a whole open ball of counterexamples. To see that, assume a counterexample simulation $x(t)$ for $t \in [0, T]$ that is assigned to a demonstration $(\tilde{x},\tilde{u})$ and assume time instants $\theta$ and a certificate candidate $L_p$. Since, $x(t)$ is a counterexample, $x(t) \notin G$ for all $t \in [0, T]$ and also for all time instants $\theta$ either $x(\theta_i) \notin I\cup G$ or $L_p(x(\theta_{i+1})) - L_p(x(\theta_i)) > \gamma[L_p(\tilde{x}(\theta_{i+1})) - L_p(\tilde{x}(\theta_i))]$. 

Due to the fact that trajectory $x(t)$ is compact, there is a time-varying neighborhood $H(t)$ around $x(t)$ for all $t\in [0,T]$ such that system simulations, that lie in $H(t)$, are also counterexample simulations. This observation in combination with continuity of ODE solutions wrt. to initial conditions and Assumption (\,IV\,) implies that there is indeed a whole open ball of counterexamples.

\section{Proof of Proposition 3}

\begin{customProposition}
Assume a finite set of demonstrations~$\mathcal{T}$ provided by a continuous and consistent demonstrator and an~assignment rule that is dominated by the~Euclidean norm. Let
$\USwitch{\mathcal{T}}{\mathcal{U}}$ be a switching controller based on $u_{\mathrm{track}}$ that meets Assumption (\,V\,), and let $(\tilde{x},\tilde{u})$ be a demonstration. Then for all $t\in [0, T]$, 
$$ \lim_{x_0 \rightarrow \tilde{x}(0)}\left|\Cost{V_t}{x_0}{\USwitch{\mathcal{T}}{\mathcal{U}}} - \Cost{V_t}{\tilde{x}(0)}{\tilde{u}}\right| = 0.$$
\end{customProposition}

\begin{proof}
Assume a demonstration $(\tilde{x},\tilde{u})$ and a time instant $t_1 \in [0, T]$. Due to consistency, a demonstration $(\tilde{x}_1,\tilde{u}_1)$ starting from $\tilde{x}_1(0) \equiv\tilde{x}(t_1)$ meets $ (\tilde{x}(t),\tilde{u}(t)) = (\tilde{x}_1(t - t_1),\tilde{u}_1(t-t_1))$ for all $t\in[t_1,T],$ i.e. segments of demonstrations starting from the same state are identical. With this observation in mind, we are ready prove the proposition itself. 

Consider a switching~trajectory $(x^\mathcal{T}_\mathrm{switch}, \LQRSwitch{\mathcal{T}})$ with $M$ switches that~ends with a~state $x^\mathcal{T}_\mathrm{switch}(T)$ in an~$\varepsilon_1$-neighborhood of $\tilde{x}(t)$. Notice that~$M$ is necessarily bounded since $\theta$ is finite. Let $t_M$ be the time instant of the~last switch. Due to continuity of the~tracking trajectory and domination of the~assignment rule, there~is an $\varepsilon_2>0$ such that all switching trajectories $(x^\mathcal{T}_\mathrm{switch}, \LQRSwitch{\mathcal{T}})$ with a state in the $\varepsilon_2$-neighborhood of $\tilde{x}(t_N)$ in time $t_M$ meet
$$\norm{(x^\mathcal{T}_\mathrm{switch}(\tau), \LQRSwitch{\mathcal{T}}(\tau)) - (\tilde{x}(\tau), \tilde{u}(\tau))} <  \varepsilon_1$$
for all $\tau\in\left[t_M, T\right]$, since the demonstrator is continuous and consistent. We will repeat the same argument for the~second last switch $t_{N-1}$ and find an $\varepsilon_3>0$ and continue on. In the~end, we will find an~$\varepsilon_{N+1}$-neighborhood of $\tilde{x}(0)$ such that all switching trajectories with an initial state in this neighborhood for given switching scheme  
$$\norm{(x^\mathcal{T}_\mathrm{switch}(\tau), \LQRSwitch{\mathcal{T}}(\tau) - (\tilde{x}(\tau), \tilde{u}(\tau))} <  \varepsilon_{N+1}$$
for all $\tau\in\left[0, t_1\right]$. Hence, any switching trajectory $(x^\mathcal{T}_\mathrm{switch}, \LQRSwitch{\mathcal{T}})$ with an initial state in this $\varepsilon_{N+1}$-neighborhood meet
\begin{equation*}
\norm{(x^\mathcal{T}_\mathrm{switch}(\tau), \LQRSwitch{\mathcal{T}}(\tau)) - (\tilde{x}(\tau), \tilde{u}(\tau))} <  \varepsilon_1 = \max_i \varepsilon_i
\end{equation*}
for all $\tau\in\left[0, t\right]$. Consequently we can make the~difference in cost for any given switching scheme arbitrarily small as well. 
\end{proof}

\section{Proof of Lemma 1}
\begin{customLemma}
For any $\varepsilon > 0$ there is a $\delta >0$ such that for any $x_0\in\mathbb{R}^n$ and for any demonstration $(\tilde{x},\tilde{u})$ with $\norm{x_0 - \tilde{x}(0)}< \delta$ and for all $t\in[0, T]$,
$$\norm{\Sol{x_0}{\LQRTrack{(\tilde{x},\tilde{u})}}(t) - \tilde{x}(t)} < \varepsilon.$$
\end{customLemma}
\begin{proof} We split the proof into two parts. In the first part, we will note some well known properties of funnels~\cite{SOS_funnels}. The second part covers the actual proof itself that follows the proof of Lemma~4 of Tobenkin~et.~al.~\cite{SOS_funnels}.

\textbf{1.}  Assume a positive definite differentiable function $P(x,t)$ and differentiable function $\rho(t)$. A time-varying set of states $\mathcal{F}(t) = \{x \in\mathbb{R}^n \mid P(x-\tilde{x}(t),t) \leq \rho(t) \}$ is a funnel provided that
\begin{equation}
\label{proof_funnel_cond}
\dot{P}(x-\tilde{x}(t),t) < \dot{\rho}(t)
\end{equation}
on the set $\{x \in\mathbb{R}^n \mid P(x-\tilde{x}(t),t) = \rho(t) \}$ for all $t\in[0, T]$. Due to \eqref{proof_funnel_cond}, any evolution that starts in the funnel $\mathcal{F}$ stays inside the funnel.  

To prove the lemma, we will construct a certain funnel around the demonstration $(\tilde{x},\tilde{u})$. Let $S(t)$ be the solution of Riccati equation \eqref{proof_riccati}. We denote $V^{(\tilde{x},\tilde{u})}(x,t) = (x-\tilde{x}(t))^TS(t)(x-\tilde{x}(t))$. First, for a given $\varepsilon>0$, we pick a constant $\rho>0$ such that  $\{x \in\mathbb{R}^n \mid V^{(\tilde{x},\tilde{u})}(x,t) \leq \rho\} \subset \{x \in\mathbb{R}^n \mid (x-\tilde{x}(t))^T(x-\tilde{x}(t)) < \varepsilon\}$ for all $t\in[0, T]$. Such a $\rho>0$ necessarily exists due to~\eqref{A5}. This set is not a funnel in general but we can consider the rescaling~\cite{SOS_funnels} 
\begin{equation}
\label{proof_funnel_cert}
 V^{(\tilde{x},\tilde{u})}_k(x,t) = e^{k\frac{T-t}{T}} (x-\tilde{x}(t))^T S(t) (x-\tilde{x}(t))
\end{equation}
parametrized by $k>0$. By increasing the value of $k,$ we reduce the size of sets $\mathcal{F}_k(t) = \{x \in \mathbb{R}^n \mid V^{(\tilde{x},\tilde{u})}_k(x,t) \leq \rho\}$. We will show that there is a value of $k$ large enough that makes  $\mathcal{F}(t)$ a funnel.

\textbf{2.}  Consider the system $\dot{x} = F(x,\LQRTrack{(\tilde{x},\tilde{u})})$  that uses the LQR tracking controller. We compute the Taylor expansion with center $\tilde{x}$ of the right hand side of this differential equation 
\begin{equation}
\label{proof_taylor}
\Delta \dot{x} = \tilde{A}(t)\Delta x + f(x,\tilde{x},\tilde{u},K(t),t),
\end{equation}
where $\tilde{A}(t) = A(t) - B(t)K(t)$ and $A(t) = \frac{\partial F}{\partial x} (\tilde{x}(t),\tilde{u}(t))$, $B(t) = \frac{\partial F}{\partial u} (\tilde{x}(t),\tilde{u}(t))$. The remainder in the Taylor formula $f = (f_1, \ldots f_n)$ has elements 
\begin{equation}
f_{i}(x,\tilde{x},\tilde{u},K(t),t) =  \Delta x^T\nabla^2 F_i(y(t),\LQRTrack{(\tilde{x},\tilde{u})}) \Delta x,
\end{equation}
where $y(t) = \tilde{x}(t) + c(t)\Delta x(t)$ for some $c(t)\in[0,1]$. 

Since $F$ is smooth, and $(\tilde{x},\tilde{u})$ is bounded, and $K(t) = -R^{-1}B(t)^TS(t)$ is bounded due to \eqref{A5}, there is an $\alpha>0$ such that for sufficiently small $\Delta x$, $|f_i| \leq \alpha \norm{\Delta x}^2$ for all $i = 1, \ldots, n$. 

Next, we compute the derivative of $V_k$ \eqref{proof_funnel_cert} along the dynamics $F(x,\LQRTrack{(\tilde{x},\tilde{u})})$ using expansion \eqref{proof_taylor}
\begin{multline}
\dot{V}^{(\tilde{x},\tilde{u})}_k(\Delta x,t) = e^{k\frac{T-t}{T}}\left(2\Delta x^TSf - \Delta x^T\left(Q + K^TRK + kS\right)\Delta x\right). 
\end{multline}
Now, notice that the derivative of $V^{(\tilde{x},\tilde{u})}_k$ is negative for all $t\in[0, T]$ and for sufficiently small $\norm{\Delta x}$ provided that $k$ is large enough. The reason is that  
\begin{equation}
\left|2\Delta x^T Sf \right|  <   \Delta x^T  (Q + K^TRK + kS) \Delta x 
\end{equation}
for sufficiently small $\Delta x$ and $k$ large enough due to our choice of $\Delta x$ and the fact that $S \geq c_1I$ due to~\eqref{A5}.

Hence, $\mathcal{F}_k$ is a valid funnel for large enough $k$. Additionally, any state that meets the inequality
\begin{equation}
\rho \geq \left| V^{(\tilde{x},\tilde{u})}_k(x,t) \right| \geq c_1 e^{k} \norm{x - \tilde{x}(t)}^2.
\end{equation}
for a given $t\in[0, T]$ belongs to $\mathcal{F}(t)$. Consequently, all states that meet 
$\norm{x - \tilde{x}(t)}^2 \leq \frac{\rho}{c_1 e^{k}} $ stay in the funnel $\mathcal{F}_k(t)$, and thus $\norm{\Sol{x}{\LQRTrack{(\tilde{x},\tilde{u})}}(t) - \tilde{x}(t)} < \varepsilon$ for all $t\in [0,T]$ due to our choice of $\rho$, which is what we have wanted to prove. 

All that remains to show is the fact that this found neighbourhood with $\delta = \sqrt{\frac{\rho}{c_1 e^{k}}}$ is valid for any demonstration. Notice however that the funnel was constructed using upper estimates of $\norm{f}$ and $V^{(\tilde{x},\tilde{u})}(x,t)$ over all possible demonstrations, i.e., independently of a particular demonstration. This observation completes the proof.
\end{proof}

\section{Proof of Lemma 2}

\begin{customLemma}
Assignment rule~\eqref{LQRarule} with any resolution of ties is dominated by the Euclidean norm, and meets assumption (\,IV\,) for some appropriate resolution of ties.
\end{customLemma}
\begin{proof} Let $\mathcal{T} = \{(\tilde{x}_1,\tilde{u}_1), \ldots, (\tilde{x}_N,\tilde{u}_N) \}$ and let $V_i(x) = (x-\tilde{x}_i(0))^TS_i(0)(x-\tilde{x}_i(0))$, where $S_i$ is a solution to the appropriate Riccati equation. We assume assignment rule~\eqref{LQRarule} $\phi(x,\mathcal{T}) = \mathrm{arg\,min}_i V_i(x)$. We first show that $\phi$ is dominated by the Euclidean norm. 

Let $\alpha >0$, $\mathcal{T} = \{(\tilde{x}_1,\tilde{u}_1), \ldots, (\tilde{x}_N,\tilde{u}_N) \}$ and let us denote the demonstration assigned to a state $x$ by $\tilde{x}_x$. We know due to \eqref{A5} that 
\begin{equation}
c_1\norm{x-\tilde{x}_i(0)}^2 \leq V_i(x) \leq c_2\norm{x-\tilde{x}_i(0)}^2
\end{equation}
for all $i = 1, \ldots, N$. Hence, if $\norm{x-\tilde{x}_i(0)}<\sqrt{\frac{c_1}{c_2}}\alpha = \beta$ for some $i$, necessarily  $V_x(x) \leq V_i(x) \leq c_2\norm{x-\tilde{x}_i(0)}^2 < c_1\alpha^2$,
and consequently $\alpha > \norm{x-\tilde{x}_x(0)}$.

Next, we want show that $\phi$ meet Assumption (\,IV\,) with a proper resolution of ties. We first notice that Assumption (\,IV\,) necessarily holds for all $x$ in which $V_i(x) < V_j(x)$ for all $j\in\{1, \ldots, N\}$ such that $j \neq i$ due to continuity of all $V_j$s. Hence, we are only interested in ties. Let us assume such a tie and without loss of generality, let us shift $x$ to the origin. The state $x = 0$ meets (we will write further $\tilde{x} = \tilde{x}(0)$ and $S = S(0)$ for short)
\begin{equation}
\label{value_proof}  
\tilde{x}_{i_1}^TS_{i_1}\tilde{x}_{i_1}  = \cdots = \tilde{x}_{i_K}^TS_{i_K}\tilde{x}_{i_K}
\end{equation}
where the tie is in indices $K = \{i_1, \ldots, i_K\}$. Now, we want to show that in any neighborhood of $x$, there is an open ball only containing states that are all assigned to one index $k \in K$.

First, we find an index $k_\mathrm{max}\in K$ such that $\norm{\nabla V_{k_\mathrm{max}}(0)}$ is maximal on $K$. If there is no other index in $K$ with the same norm of gradient at $0$, we are done, since there would be $y$ in the direction of $\nabla V_{k_\mathrm{max}}(0)$ from $0$ such that $V_{k_\mathrm{max}}(y) < V_k(y)$ for all $k \neq k_\mathrm{max}$. Using the same reasoning, there must be other indices $l\neq k_\mathrm{max} \in K$ such that these gradients are equal to $\nabla V_{k_\mathrm{max}}(0)$, i.e.,
\begin{equation}
\label{grad1_proof} 
\nabla V_{k_\mathrm{max}}(0) = -2S_{k_\mathrm{max}}\tilde{x}_{k_\mathrm{max}} = \nabla V_l(0)
\end{equation}
otherwise, we would be done as well. Let us denote set of all such indices by $L$. Then there is $y$ in any given neighborhood of $0$ in the direction of $\nabla V_{k_\mathrm{max}}(0)$ such that $\forall l \in L$ is $V_l(y) < V_j(0)$ for all $j\in K\setminus L.$ Due to continuity, this inequality holds on a whole neighborhood of $y$ which we denote $H$. 

Now, let us assume that this neighborhood does not contain an open ball of states that are assigned to the same demonstration. This is possible only when there is a tie for all $z\in H$. Let us split all states $z\in H$ into finitely many categories according to in which indices the tie is and which gradients are the greatest in terms of the norm. Since the states $z$ covers the whole neighbourhood $H$ and the number of categories is finite, there must be a category that contains $n$ (the dimension of the state space) linearly independent states $w\in H$. Otherwise the states from all categories could not cover the whole $H$.  Hence, there must be two indices $k_1$ and $k_2$ such that $S_{k_1}(w-\tilde{x}_{k_1}) = S_{k_2}(w-\tilde{x}_{k_2})$, where $w$ are $n$ linearly independent vectors from the category corresponding to the tie $V_{k_1}(w) = V_{k_2}(w)$ with those two having the greatest norm of gradient. Thus, $S_{k_1}$ must equal $S_{k_2}$. Which in turn implies from \eqref{grad1_proof} that $\tilde{x}_{k_1} = \tilde{x}_{k_2}$. Hence, there cannot be a tie for all $z\in H$, which completes the proof. 
\end{proof}

\end{document}